\documentclass[a4paper]{article}

\usepackage{amsmath,amssymb}

\usepackage{changepage}

\usepackage[utf8x]{inputenc}

\usepackage{textcomp,marvosym}

\usepackage{cite}

\usepackage{microtype}
\DisableLigatures[f]{encoding = *, family = * }

\usepackage[table]{xcolor}

\usepackage{array}

\usepackage[T1]{fontenc}% optional T1 font encoding
\usepackage{graphicx}% for including images
\usepackage{float}% for using [H] after including image to place it
\usepackage{multirow}% for mutlirows in tables
\usepackage{amsmath} % *** MATH PACKAGES ***
\usepackage[cmintegrals]{newtxmath}
\usepackage{color,soul} % for higlighting adjustments
\usepackage{cite}
\usepackage{threeparttable} %For note in table
\usepackage{nameref, hyperref}
\usepackage{subcaption}
\usepackage[export]{adjustbox}
\usepackage{array}
\usepackage{capt-of}% or \usepackage{caption}

\newcommand{\scantonum}{{\sc scan2num}}
\newcommand{\note}[1] {
  \textcolor{red}{Note: #1}}
\renewcommand{\note}[1]{}

\title{Function Follows Form: Regression from Complete Thoracic Computed Tomography Scans}
\date{}

\begin{document}
\maketitle
%\vspace*{0.2in}

\begin{flushleft}
%{\Large
%\textbf\newline{Function Follows Form: Regression from Complete Thoracic Computed Tomography Scans} % %Please use "sentence case" for title and headings (capitalize only the first word in a title (or %heading), the first word in a subtitle (or subheading), and any proper nouns).
%}
%\newline
%\\
Max~Argus\textsuperscript{1},
Cornelia Schaefer-Prokop,\textsuperscript{2},
David A. Lynch\textsuperscript{3}, 
Bram~van~Ginneken\textsuperscript{1},
\\
\bigskip
\textbf{1} Diagnostic Image Analysis Group at Radboud University Medical Center, Nijmegen, The Netherlands.
\\
\textbf{2} Meander Medical Center, Amersfoort, The Netherlands.
\\
\textbf{3} Jewish National Health, Denver, CO., USA.
\\
\bigskip

% Use the asterisk to denote corresponding authorship and provide email address in note below.
%* correspondingauthor@institute.edu

\end{flushleft}
\section*{Abstract}
Chronic Obstructive Pulmonary Disease (COPD) is a leading cause of morbidity and mortality. While COPD diagnosis is based on lung function tests, early stages and progression of different aspects of the disease can be visible and quantitatively assessed on computed tomography (CT) scans. Many studies have been published that quantify imaging biomarkers related to COPD. In this paper we present a convolutional neural network that directly computes visual emphysema scores and predicts the outcome of lung function tests for 195 CT scans from the COPDGene study. Contrary to previous work, the proposed method does not encode any specific prior knowledge about what to quantify, but it is trained end-to-end with a set of 1424 CT scans for which the output parameters were available. The network provided state-of-the-art results for these tasks: Visual emphysema scores are comparable to those assessed by trained human observers; COPD diagnosis from estimated lung function reaches an area under the ROC curve of 0.94, outperforming prior art. The method is easily generalizable to other situations where information from whole scans needs to be summarized in single quantities.

% Please keep the Author Summary between 150 and 200 words
% Use first person. PLOS ONE authors please skip this step. 
% Author Summary not valid for PLOS ONE submissions.   
%\section*{Author summary}
%Lorem ipsum dolor sit amet, consectetur adipiscing elit. Curabitur eget porta erat. Morbi consectetur est vel gravida pretium. Suspendisse ut dui eu ante cursus gravida non sed sem. Nullam sapien tellus, commodo id velit id, eleifend volutpat quam. Phasellus mauris velit, dapibus finibus elementum vel, pulvinar non tellus. Nunc pellentesque pretium diam, quis maximus dolor faucibus id. Nunc convallis sodales ante, ut ullamcorper est egestas vitae. Nam sit amet enim ultrices, ultrices elit pulvinar, volutpat risus.

% Use "Eq" instead of "Equation" for equation citations.
\section{Introduction}
Volumetric medical scans contain a wealth of information, spatially encoded in a very large number of voxels, in the order of $10^9$ for computed tomography (CT) scans. Often clinicians are interested in single quantities that represent the status of the patient. Ideally, CT scans would allow such measurements in an objective and repeatable manner. From a machine learning perspective, reducing $10^9$ measurements to a single number is a daunting task. As a consequence, automated systems that have been developed to carry out such tasks typically employ a sequence of steps: data is first preprocessed, structures that are considered most relevant are segmented, within these structures certain regions are identified where specific measurements are performed, and these measurements are aggregated into the final quantity to be extracted. Clearly, such systems are highly task specific and result from a large number of dedicated design decisions. 

In this work we pursue an orthogonal approach and present a system, based on a convolutional network, that is intended to generically carry out the task of translating a volumetric scan directly into a single measurement. The system is trained end-to-end using a set of pairs of scans (the input) and the quantity to be derived from the scans (the output). No specific medical knowledge about what in the scan is important to derive the desired output is encoded into the system. We therefore refer to our system with the general name \scantonum.

We apply \scantonum \ to thoracic computed tomography (CT) scans. We present results for various tasks related to chronic obstructive pulmonary disease (COPD). This is currently the fourth leading cause of death worldwide, and expected to be the third leading cause of death in 2020. In COPD patients chronic inflammation of the airways and the lungs, usually due to smoking, leads to structural changes in the airways and lung parenchyma. The airflow limitation in COPD is assessed spirometrically by a combination of forced expiratory volume in 1 second (FEV$_1$) and the forced vital capacity (FVC). From these quantities the GOLD score is computed, with which COPD is diagnosed and the severity of the disease is quantified.

Where spirometry measures global functional aspects of the disease, CT spatially resolves the pathophysiological processes and structural damage. In the CT scans airway wall thickening can be observed, as well as emphysema resulting from degradation of lung tissue through the enlargement and destruction of alveoli. CT shows both the severity and extent of different emphysematous patterns, such as centrilobular, parenchymal and paraseptal emphysema. This information is reported by radiologists to inform treatment decisions \cite{Lync15}. Visual assessment has, however, been found to have a large inter-reader variability when reporting these quantities \cite{Lync12}.
\note{Max: Are these the correct citations}

Ideally, analysis of CT scans would provide sensitive and repeatable measurements of disease progression. From CT scans it may be possible to quantify not only physical progression, but also the functional progression by training models to predict functional quantities. Such appearance based measurements could be further used for screening for COPD, for monitoring and for measuring response to treatment. Imaging biomarkers that correlate with progression are increasingly attracting attention, as a way to measure the effects of different treatment options \cite{Mosl16, Male14}.

To demonstrate the potential of \scantonum, we apply the network in this study to various quantitative tasks related to COPD. We  trained it to predict standardized parenchymal emphysema scores, and to assess the lung function test scores FEV$_1$/FVC and FEV$_1$\%predicted directly from the CT scan. These two scores can be used to diagnose COPD and determine the GOLD stage of the patient. 

We believe that this network can also be applied to other, similar, problems and hopefully is a step in the direction of more end-to-end optimization for CT.

\subsection{Related Work}

In this section we briefly review prior work on the tasks we address in this study, namely emphysema quantification and deriving estimates of lung function measurements from CT scans, and we report on related work on deriving measurements from 3D scans using deep networks. 

%This requires the aggregation of information from the whole scan into a single quantity.
%How best to best aggregate depends on the diagnosis one is trying to make.

%The task becomes more difficult if the decision cannot be made based on a local bounding box, i.e. like segmentation or nodule detection, but instead requires seeing a more global scope.
%This is especially true for CNNs as all intermediate features of the computation need to be held in memory in order to be able to perform back-propagation. 

%The second problem is memory footprint of the computation.
%While many previous algorithms, such as emphysema score computation, also map whole scans to single quantities, the ease of doing so depends mainly on how the algorithms factor spatially.

Emphysema quantification can be performed using a densitometric approach or a texture based approach.
In the standard densitometric analysis of emphysema an emphysema score is computed as the percentage of lung tissue, excluding airways, with a density less than a given threshold value \cite{Rikx09}. 
This method has been applied to the COPDGene dataset in \cite{Han11}.

As this standard method suffers from lack of reproducibility a number of correction methodologies were introduced, such as volume correction \cite{Stoe08} and kernel hardness correction \cite{Gall16}.

%To make scans compareable emphysema scores are commonly evaluated on inspiration scans, with some degree of airway masking.

%In theory these should be sufficient as Houndsfiled Units(HU) are a well defined physical quanity.
%In practice these methods suffer from a lack of reproducability due to the effect of different CT machine settings.
%To alleviate this problem the post-processing of CT scans has been explored to retroactivley normalize the effects of different CT settings
%Alternativley there has been research into texture-based methods to analyze CT scans.
%These methods promise a greater degree of reproducibility as they are less dependent on calibration.

As an alternative to purely densitometric approaches texture based approaches have also been introduced. The prior work in \cite{Xu06}, \cite{Park08},\cite{Sore10}, \cite{Haem14}, \cite{Yang16}, \cite{Gins16} usually starts by extracting local, hand crafted features from Regions of Interest(RoI). These features are then used to train a classifier to classify the RoIs into the previously annotated lung tissue types. Sometimes local features are also learned based on manual RoI annotations. 

Clustering avoids manual RoI annotation and is applied to emphysema subtyping in \cite{Bind16}, as does multiple instance learning, which is applied to the problem of COPD classification in \cite{Chep14}. This approach, nevertheless, requires hand crafted features as a basis.

Prior work in predicting lung function from CT scans also relies on hand crafted features.
These features are often quite coarse, relying on organ level thresholds, medians and volumes as can be seen in \cite{Mets12}. \note{Which Keelin Murphy paper, others?}

Transitioning from hand crafted features to learned features is an important step as the latter can often be more expressive and efficacious. Similarly not relying on RoIs is also important as, in many cases, the functional quantities of interest cannot be directly associated with visual patterns.

Scan-level CNN models have also recently been used for
%Alzheimer's Disease classification in \cite{Hoss16} based on whole MRI scans and for
longevity prediction from CT scans in \cite{Oakd17}. This approach makes use of a 3D CNN model, but therefore requires aggressive down-sampling of the scans. We instead advocate a late fusion architecture, as this allows processing the scans at full resolution which better enables texture based emphysema classification. \note{is full resolution required?}

\section{Materials and methods}
\subsection{Subjects and Image Acquisition}

A major hurdle to applying large models, such as convolutional networks, to make scan level predictions is that a large number of scan-level examples are required to avoid overfitting. 
The chest CT scans used in this study were obtained from the multi-center COPDGene Study. 
This study included 10,192 current or former smokers and 108 non-smokers. 
Details regarding the inclusion criteria and CT acquisition can be found in the COPDGene study design paper \cite{Rega10}.
Briefly, subjects were scanned at full inspiration with 200 mAs. Thin-slice reconstructions were used. Scans come from 21 centers and a wide variety of scanners were used. When multiple reconstruction kernels were available, we choose the softer kernel. 
%\note{you included Han11 here but that has no info on CT scanning and refers to Rega10. Include Han11 literature discussion as a standard emphysema quantification paper, but it is used on copdgene so relevant}. \cite{Han11}

We used baseline CT scans from those subjects for which both the visual emphysema scores and lung function measurements were available. This yielded a set of 1,788 scans, which were randomly split into a training set of 1,424 scans a validation set of 170 scans and a test set of 195 scans, stratified according to COPD Gold stages and outcome parameters. 

%To improve this situation a randomly sampled set of COPDGene cases was chosen.
%VE scores were only available for 3,146 randomly selected patients.

\subsection{Assessment of Visual Emphysema Score}

The large amount of inter-reader variability in emphysema classification and quantification has lead to efforts to precisely delineate different visual phenotypes \cite{Lync15} and create reader training and scoring procedures in order to produce consistent measures.

This was done for parenchymal emphysema, the most common type of emphysema for COPD patients. 
In parenchymal emphysema small lesions form at the center of the secondary lobule and in case of progression they coalesce into large areas of lung destruction.
%In PSE small legions form near the pleural surface close to the chest wall and develop similarly.
The visual scoring of emphysema follows the methodology of \cite{Halp17}. The severity of parenchymal emphysema was assessed by trained medical analysts according to a six point scale of: no emphysema, trace,  mild, moderate, confluent or advanced destructive emphysema. The analysts scored the highest grade of emphysema that is visible on the scans, even if it was present only in one location. Weighted kappa scores for observer agreement for this scoring system ranged from 0.70 to 0.80, indicating good to excellent agreement. This quantity is referred to as the Visual Emphysema (VE) score.

%This quantity is referred to as the Visual Emphysema (VE) score, as the more advanced stages are not CLE.

%Visual scoring of emphysema by experienced observers according to a standardized methodology can provide incremental information over and above standard quantitative measures.

%This high inter-observer variability is an inherent problem in lung disease analysis, and indicated that a informative and consistent method to do automatic measurement would be valuable.

%These scores were used to train and evaluate an automatic scoring system.
%While visual scoring has the potential of producing more meaningful information than radiometric scoring methods it has the drawback of re-introducing subjectivity into labeling.
%The inter-observer variability thus represents an (soft) upper bound on the potential performance of a labeling system.
%In the context of machine-learning it can be described as label noise.
%In a simplification we can consider n observers as independent measurements of the underlying quantity, this way we should get a rate by which the labels are wrong.

%Paraceptal Emphysema was also visually scored on the CT scans.
%The severity of PSE was assessed according to a three point scale.
%This form was not considered as extensively due to the smaller number of categories, being given for this disease.
%For many cases the PSE score 1 was given to scans in which only a very small number of slices showed the disease.
%This makes it a much harder learning problem than the more diffuse CLE.

\section{Method: \scantonum~Network}
\note{Double check it does not repeat too much what's in the introduction}
The diagnostic process aims to find a single quantity that describes the state of the patient.
Computationally this corresponds to the conversion of a spatially resolved image into a per-scan value.
This computation is often factored into a local computation plus a global aggregation step.
Many tasks such as lesion segmentation and nodule detection are focal, in the sense that decisions can be made based only on a local volume, whereas other tasks are more diffuse in nature.

While the spatial factoring of an algorithm is dictated by intrinsic attributes of the problem in focal tasks, choosing a spatial factoring for diffuse tasks is more difficult.
This is especially true when applying CNNs which need to reconcile the desire for gradual spatial resolution reduction with algorithm memory usage as for training as all intermediate features of the computation need to be held in memory in order to be able to perform back-propagation.

An extreme example of spatial factoring can be seen in emphysema score computation, which also maps whole scans to single quantities, it makes local single-voxel decision by thresholding and then aggregates these by averaging.

In this work we formulate a CNN network, named \scantonum, that addresses this problem for three important quantities in COPD diagnosis that describe lung function and aspect. 

The architecture that we chose is based on the computer vision task of video classification 
\cite{Karp14}, in which a number of architectures have been developed to solve this similar problem. For our task we chose a late-fusion architecture as these provide good classification performance. In these architectures, the same 2D network is applied to a regularly spaced subset of slices and the resulting feature vectors are averaged to aggregate the 2D slice information over the whole volume.

\begin{figure}[!t]
  \centering
  \includegraphics[width=.66\linewidth]{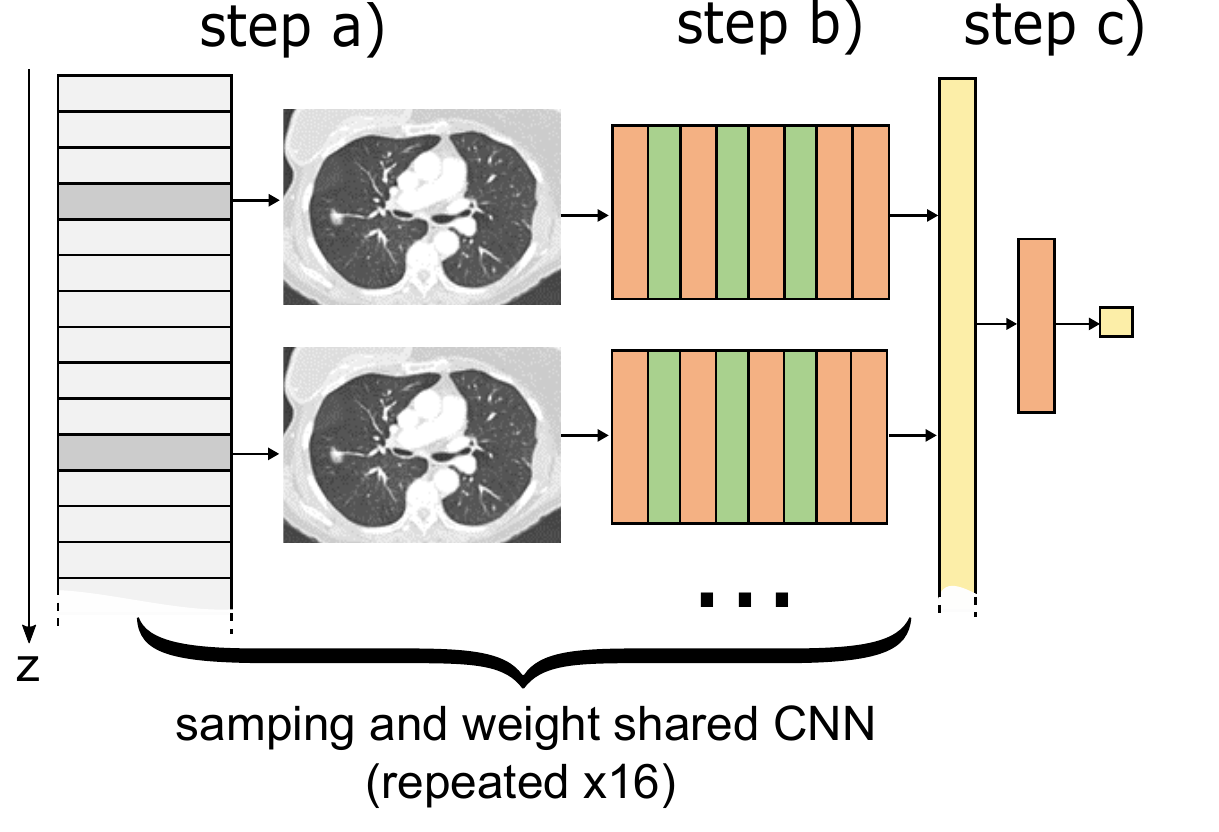}
  \caption{Late fusion \scantonum network for whole scan classification. The network follows three steps a) axial slice sampling, b) per-slice feature computation c) feature aggregation by summation and scoring.}
  \label{fig:scan2num_model}
\end{figure}%

With this choice of model the process can be described as consisting of three main steps of preprocessing, 2D per-slice feature computation and finally feature aggregation and scoring.

In pre-processing the CT scans are cropped to the bounding box of the lung segmentation. The lung segmentations were provided in the COPDGene database and were computed using a previously published approach \cite{Slui05a}. This step removes some of the irrelevant areas from the CT scans to help the remaining steps focus on the lung. Subsequently, regularly spaced slices are sampled from the scan.

In the per-slice computation the aim was to extract an informative feature vector, without letting the model become too large. 
This was done by a very narrow network in order to save memory space.
As identification of emphysema depends on the local texture of the lung, slices where given to the network at their full resolution. 

In the final step the per-slice features are aggregated by mean pooling of the per-slice feature vectors.
This allows varying the number of slices that are considered for the score.
16 slices per scan were chosen, as this was shown to produce good results see \autoref{tag:model_variation} for details.

In the scoring step the global feature vector, representing the entire volume, is converted into a per-scan score. 
A L2 regression loss was chosen as need to predict continuous quantities, and our discrete VE scores were ranked. 
This method of training has the benefit of transmuting discrete labels into continuous ones, which may be more useful for further statistical analysis.
Network architecture and training parameters can be found in \autoref{tab:arch}.

The network chosen is inspired by the second place entry in the Kaggle Diabetic Retinopathy Detection Challenge and is described in \autoref{tab:arch}.

\begin{table}[!t]
\centering
\caption{
  Network architecture. The network processes 16 slices which are aggregated into a single scan label.
  \label{tab:arch}
}

\begin{tabular}{| c | c | c  c c| c |}
\hline
step & layer name   &  channels & filter & stride & output size\\
\hline
a)& input      &  1    &     &    & 512 \\
\hline   
&conv1       &  32   & 5x5 & 2  &  254 \\
&pool1       &  32   & 2x2 & 2  &  127 \\
&conv2       &  128  & 5x5 & 2  &  62 \\
b) &pool2    &  128  & 2x2 & 2  &  31 \\
&conv3       &  256  & 3x3 & 2  &  15 \\
&pool3       &  256  & 2x2 & 2  &  8 \\
&conv4       &  512  & 3x3 & 2  &  3 \\
&conv5       &  1024 & 3x3 & 1  &  1 \\
\hline
&sum         &  1024 &     &    & 1 \\
c) &fc6      &  1024 &     &    & 1 \\
&L2 loss     &  1    &     &    & 1 \\
\hline
\end{tabular}
\end{table}

We trained our models using stochastic gradient descent with a batch size of 16 examples, momentum of 0.9.
Linear learning rate decay was used starting with a learning rate of 0.005 over 100k iterations.
The performance on the validation set was evaluated every 500 iterations and the network with the lowest loss on the validation set was chosen as the final network.
Weights are initialized using the Xavier filler \cite{Glor10}. 
Weight decay of 0.0005 was used together with dropout of 0.5 that was applied only to the fc6 layer. 
Data augmentation of random mirroring and per slice rotation by $\pm 45^{\circ}$, sampled uniformly, was applied. 
During training regularly spaced slices were sampled, by varying the inital slice offset according to a uniform distribution additional data variability was achieved.
All models are trained and tested with Caffe \cite{jia14} on a single NVIDIA GeForce GTX 1080.

% Results and Discussion can be combined.
\section{Results}
The described network was applied to compute predictions for three quantities associated with COPD, the VE score, FEV$_1$/FVC and FEV$_1$. All results are reported for the same test set of 195 scans. The comparison of performance relative to other methods is hindered by the lack of a standardized test set. \footnote{Our test set, with individual reader scores, is available here: 
\href{goo.gl/vP294U}{goo.gl/vP294U}.}
%\href{https://gist.github.com/BlGene/521e4d6d4cea44daa048e4818234f646}{here}. }
%This is important because different patient selections have different distributions of disease
%severity, which in turn affect the performance of a system.

\subsection{VE Classification}

\begin{figure*}
\centering

\begin{subfigure}[t]{.5\linewidth}
  \centering
  \includegraphics[width=\linewidth]{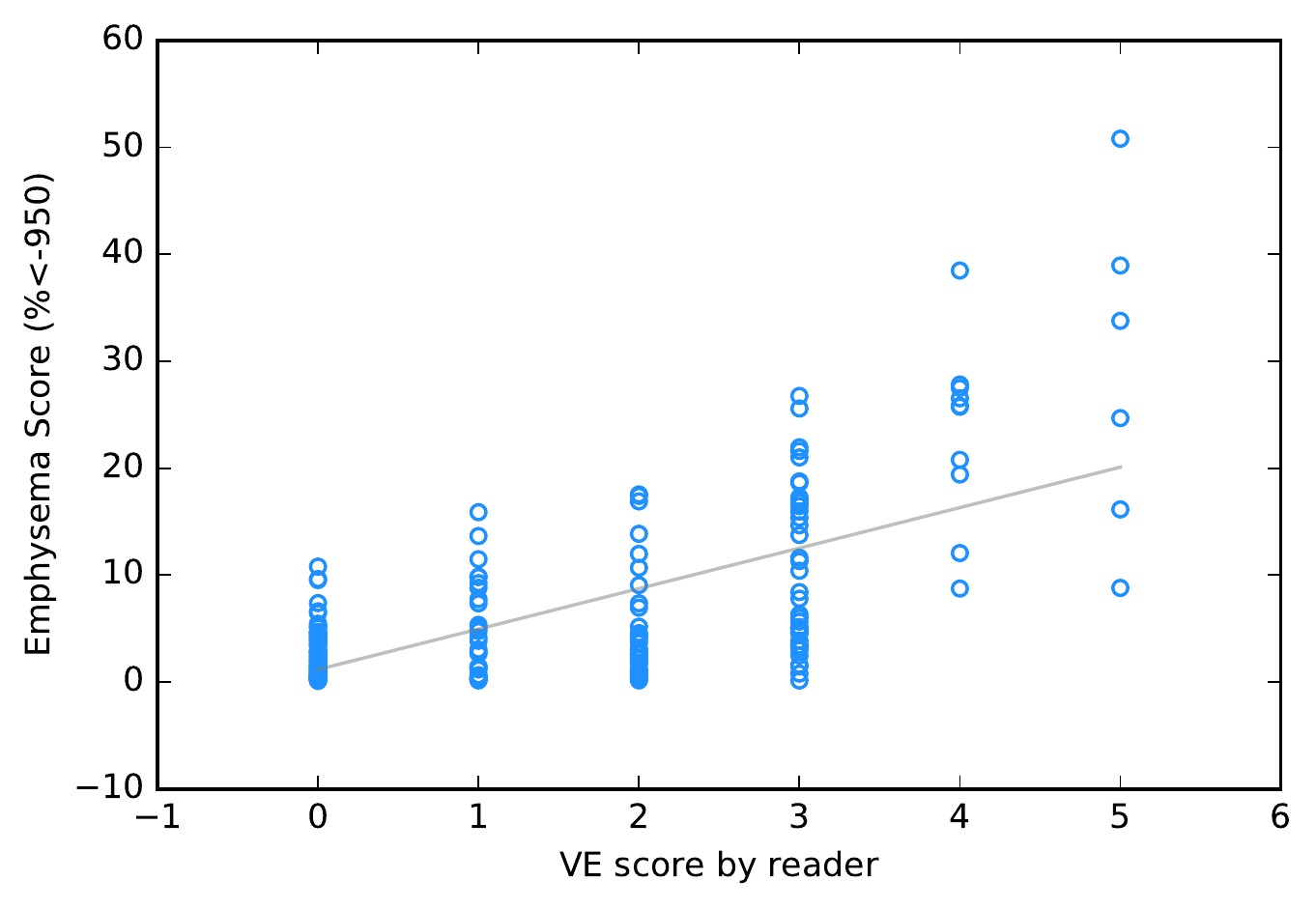}
  \caption{\%-emphysema correlation with VE \\ score  $\rho = 0.56$, 95\% CI [0.46 - 0.64]}
  \label{fig:cle_pctemph_v2}
\end{subfigure}%
\begin{subfigure}[t]{.5\linewidth}
  \centering
  \includegraphics[width=\linewidth]{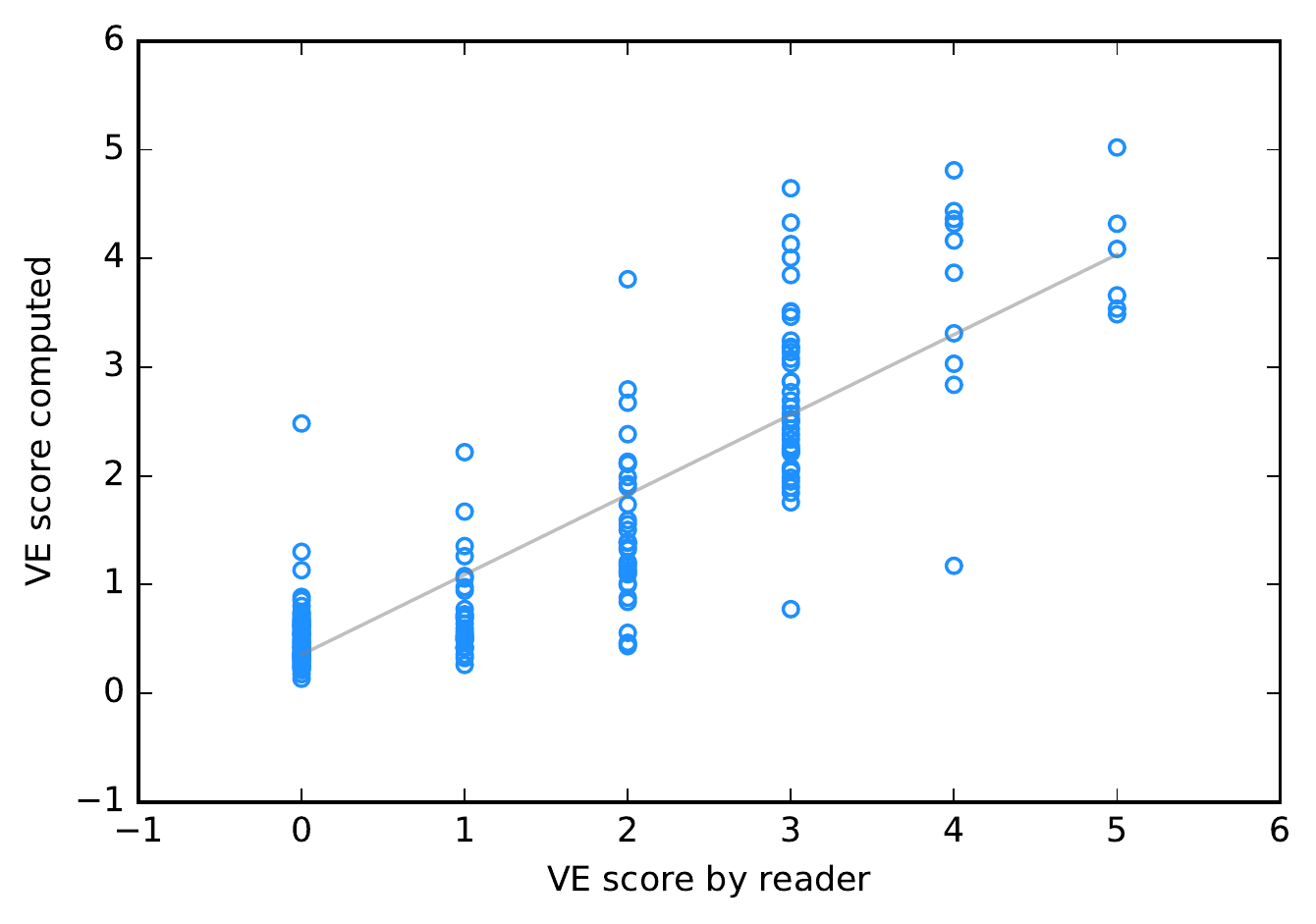}
  \caption{ VE score prediction.  $\rho = 0.84$, 95\% CI [0.80 - 0.88]}
  \label{fig:cle_predicted_v2}
\end{subfigure}

\begin{subfigure}[t]{.5\linewidth}
  \centering
  \includegraphics[width=\linewidth]{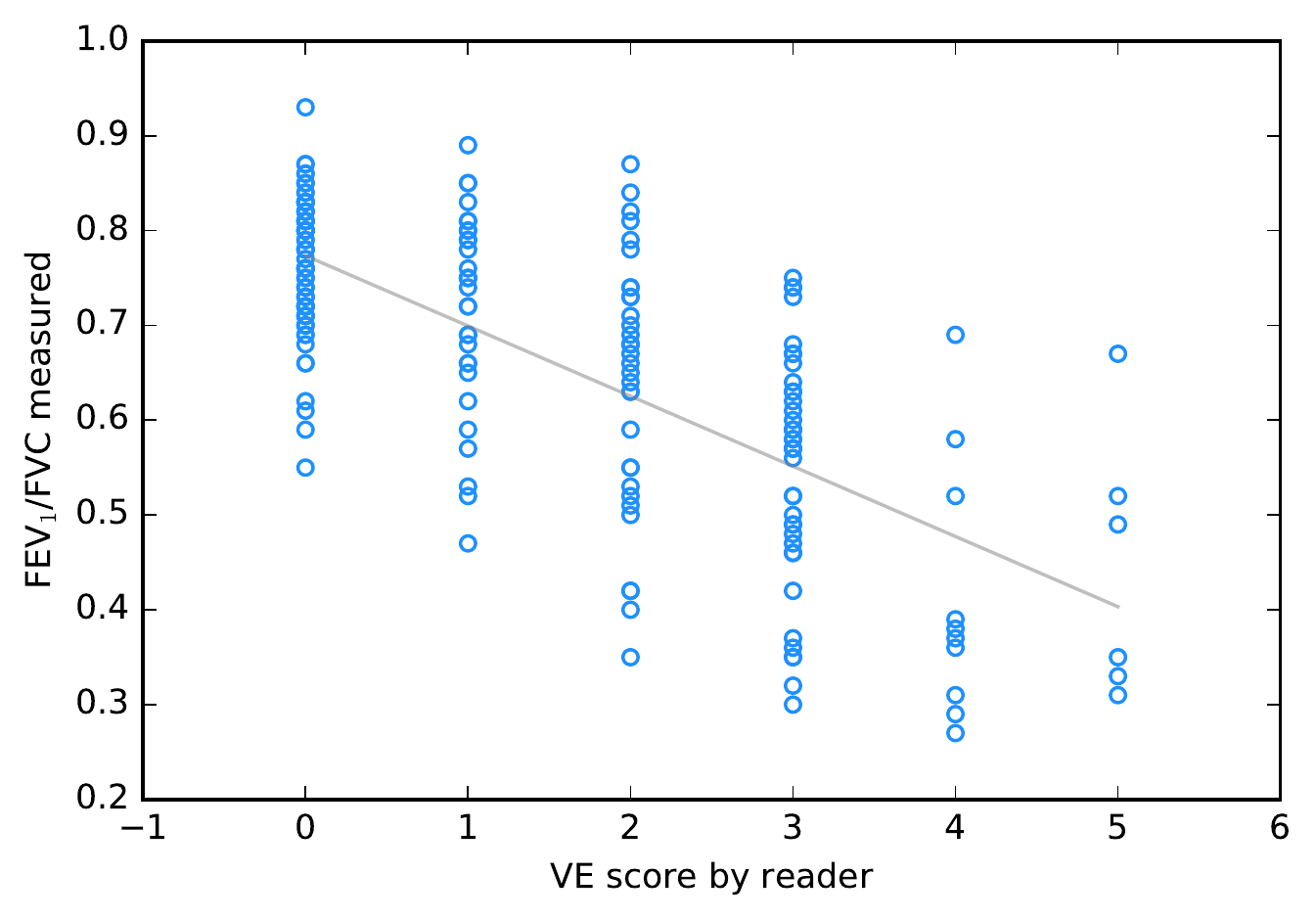}
  \caption{FEV$_1$/FVC correlation with VE  \\ score.  $|\rho| = 0.70$, 95\% CI [0.63 - 0.75]}
  \label{fig:fof_vs_cle}
\end{subfigure}%
\begin{subfigure}[t]{.5\linewidth}
  \centering
  \includegraphics[width=\linewidth]{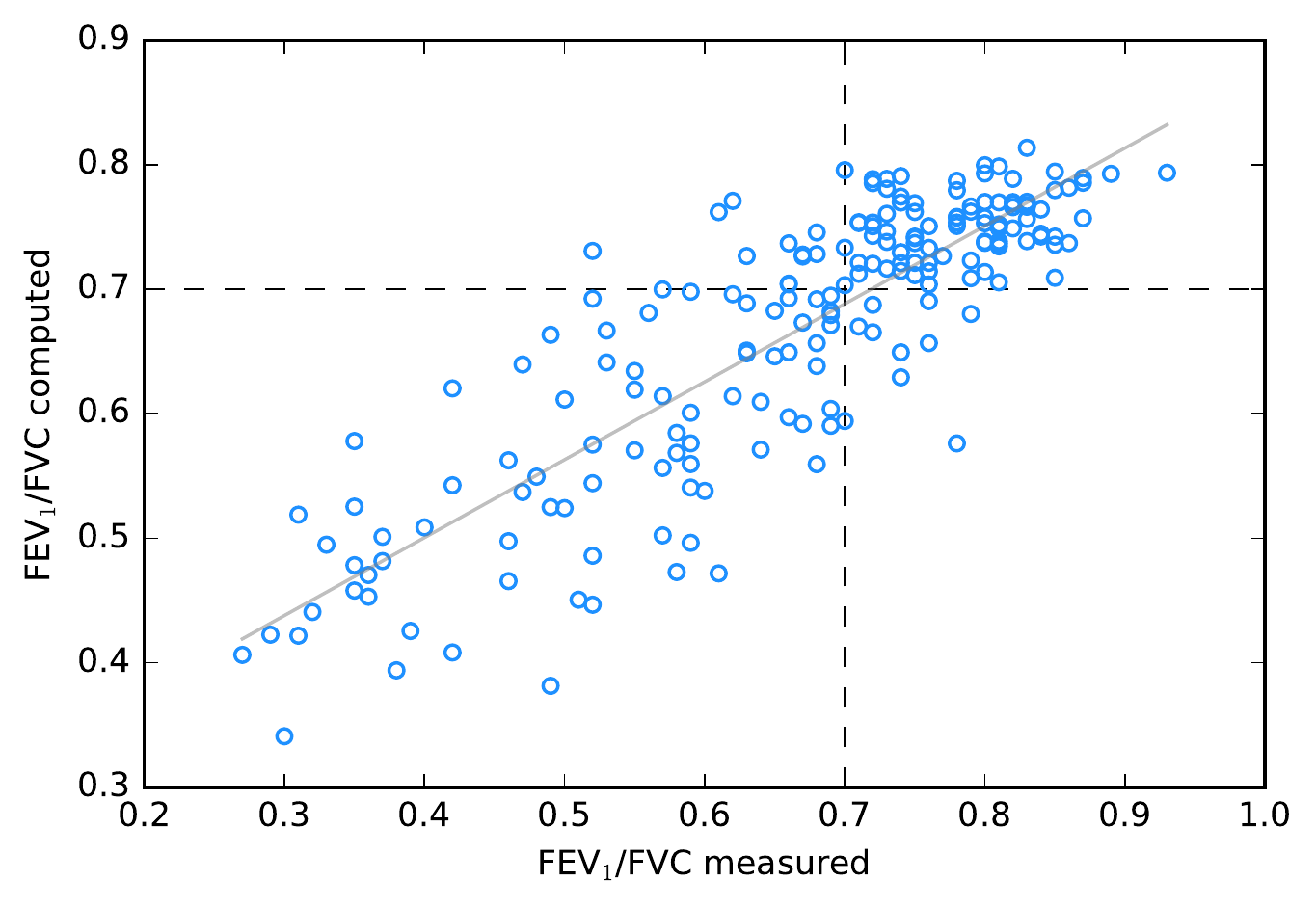}
  \caption{FEV$_1$/FVC prediction $\rho=0.82$, 95\% CI [0.77-0.86]  showing COPD threshold.}
  \label{fig:fof_computed}
\end{subfigure}

\begin{subfigure}[t]{.50\linewidth}
  \centering
  \includegraphics[width=\linewidth]{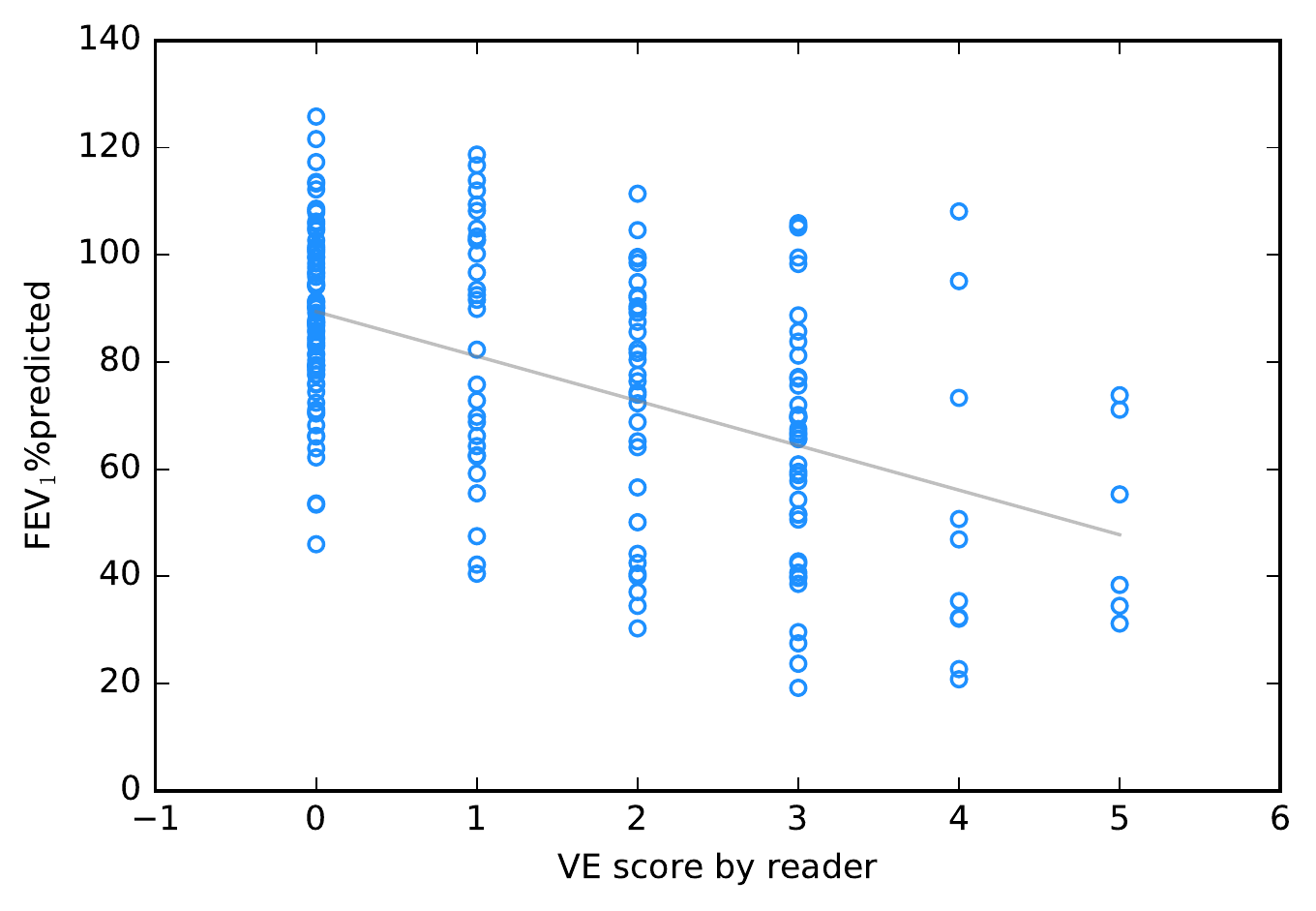}
  \caption{FEV$_1$\%predicted correlation with \\ VE score. $|\rho| = 0.47$, 95\% CI [0.37 - 0.56]}
  \label{fig:fev_vs_cle}
\end{subfigure}%
\begin{subfigure}[t]{.50\linewidth}
  \centering
  \includegraphics[width=\linewidth]{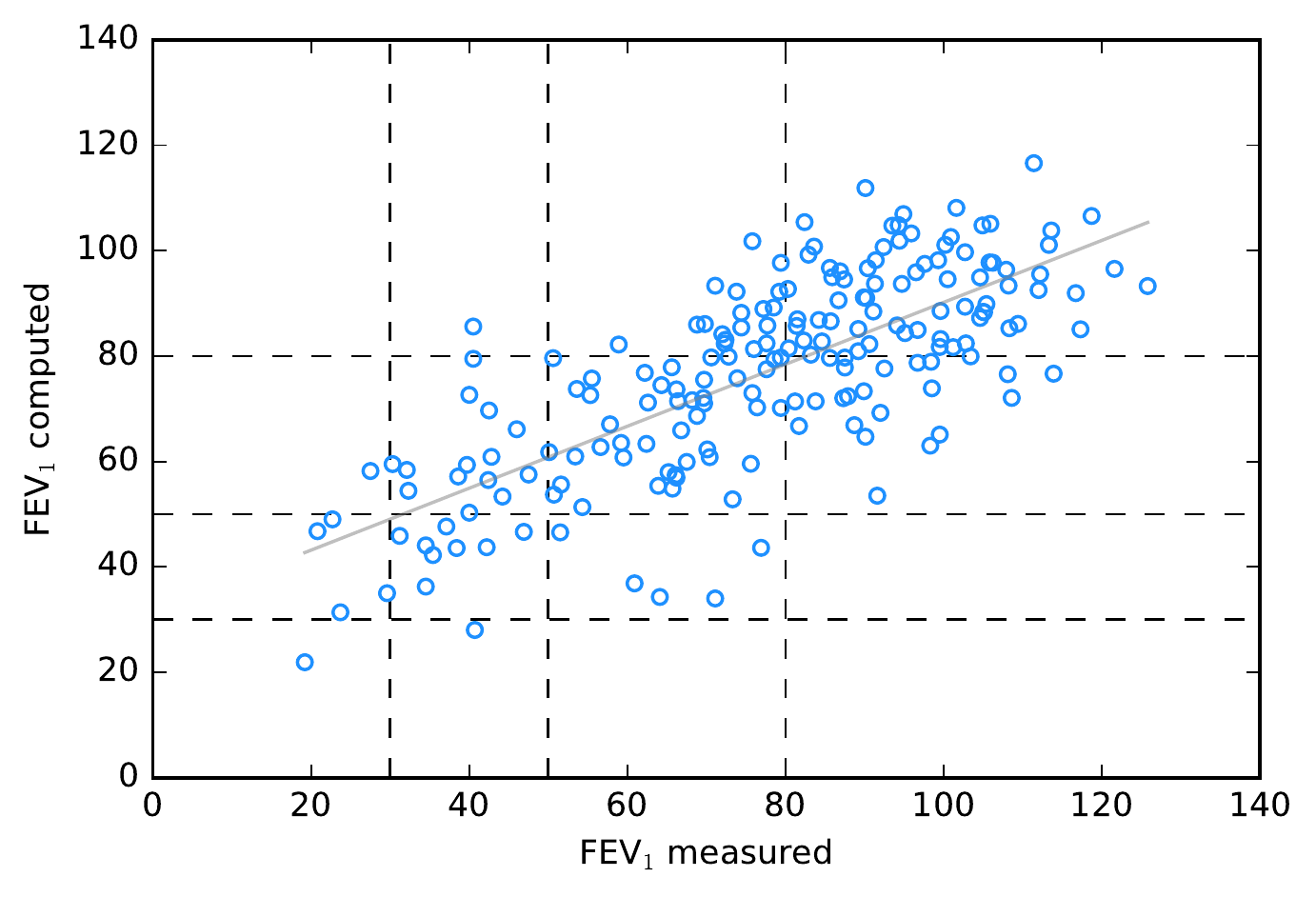}
  \caption{FEV$_1$\%predicted prediction
  $\rho = 0.73$, 95\% CI [0.66 - 0.78]
  showing GOLD stage thresholds.}
  \label{fig:fev_computed}
\end{subfigure}

\caption{Scatter plots showing the Spearman's rank correlation of computed predictions for various quantities (right column) compared to the correlation of these quantities with chosen reference values (left column). The gray lines are linear fits, all correlations shown are for the same test set of 195 patients and have $p < .001$.
\label{fig:plots}
}

\end{figure*}

\begin{figure*}[!ht]
\centering

\subcaptionbox{\label{fig:out_a}}{\includegraphics[width=.20\textwidth]{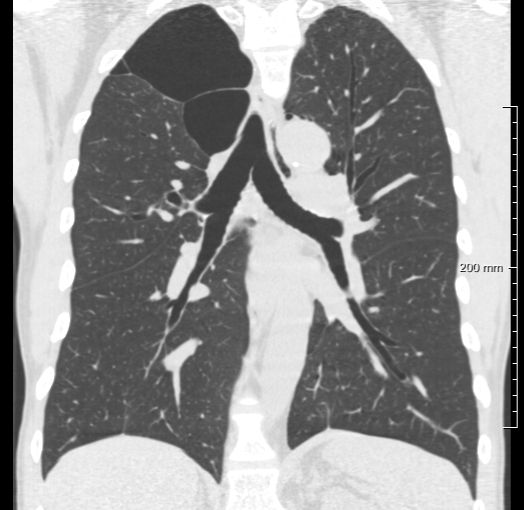}}% o
\subcaptionbox{\label{fig:out_b}}{\includegraphics[width=.20\textwidth]{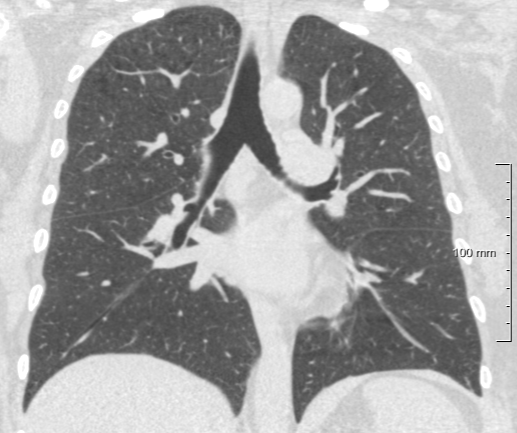}}% u
\subcaptionbox{\label{fig:out_c}}{\includegraphics[width=.20\textwidth] {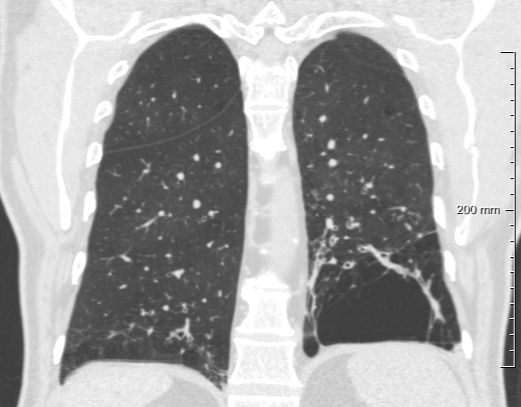}} 

\subcaptionbox{\label{fig:out_d}}{\includegraphics[width=.20\textwidth]{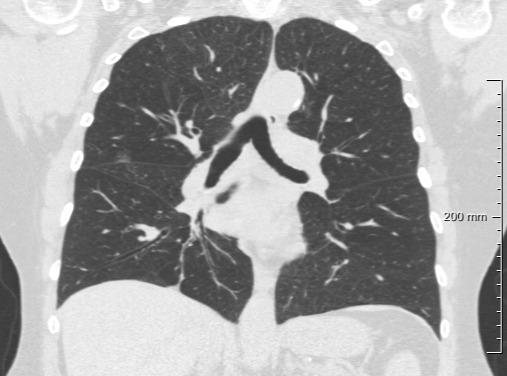}}% o
\subcaptionbox{\label{fig:out_e}}{\includegraphics[width=.20\textwidth]{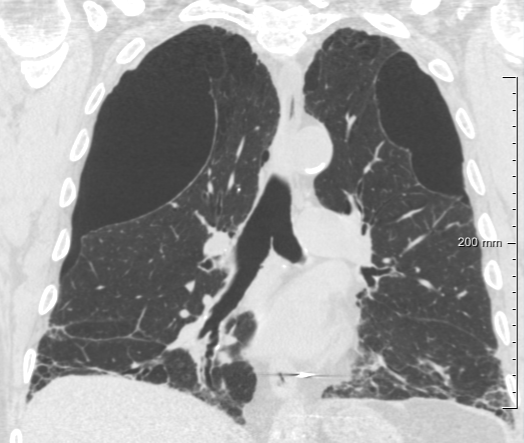}}% o
\subcaptionbox{\label{fig:out_f}}{\includegraphics[width=.20\textwidth]{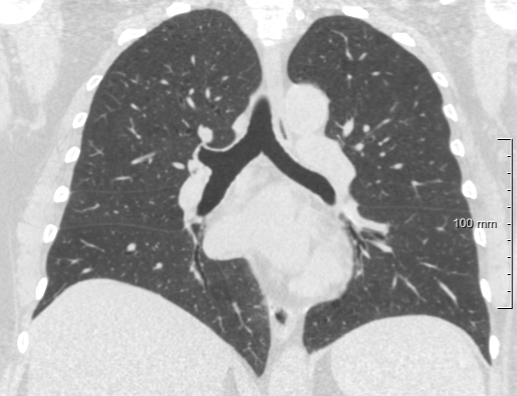}}

% a b c d e f
% o u u o o u
\caption{VE score prediction outliers. The presence of upper lobe paraseptal emphysema without parenchymal emphysema in (a)+(e) and diffuse emphysema in (d), may have lead to over-prediction, whereas lower lobe paraseptal emphysema in (c) and possible incomplete inspiration in (b)+(f) may have lead to under-prediction. \label{fig:out}}
\end{figure*}

We started with VE score prediction. As this is a quantity that is visually assessed from the scan itself, it was the most likely to be predictable.
The results are shown in \autoref{fig:cle_predicted_v2} next to a plot of the percent emphysema below the -950HU threshold as a benchmark shown in \autoref{fig:cle_pctemph_v2}. 
In the predictions we see a good separation of classes, except between VE scores of 0--1, which correspond to absent and trace amounts of emphysema. 
This may be due to the fact that small amount of emphysema are more easily missed by both the slice sampling step and the feature computation step. 
Visual assessment of outliers is shown in \autoref{fig:out}.
Further visual evaluation is possible by looking at the individual VE response per-slice. This should produce meaningful information as all feature aggregation operations are linear. The results of this can be seen in \autoref{fig:extrema}. Here we see that the per slice VE score correlates with the relative presence of emphysema.

A slight systematic offset can also be seen in the results when comparing the linear and quadratic fits of the data. These deviation between these two shows that the dataset is biased to prediction the dataset mean of $\sim 1.75$. 
For further analysis of accuracy the continuous emphysema scores were discretized.
This was done, without further calibration, by rounding to the nearest whole number. 
This results in the confusion matrix \autoref{tag:cle_cm}.
%, an accuracy of 0.43, a mean class accuracy of 0.46 and mean class Intersection over Union of 0.28. 

\begin{table}[!ht]
\centering
\caption{
  Confusion Matrix of VE scores comparing visually assessed and computed scores.
  \label{tag:cle_cm}
  }
\begin{tabular}{|c c | c c c c c c|}
\hline
 \multicolumn{2}{|c|}{VE}  & \multicolumn{6}{c|}{Predicted} \\
 \multicolumn{2}{|c|}{Score} & 0 & 1 &	2 & 3 & 4 & 5 \\
\hline
\parbox[t]{2mm}{\multirow{6}{*}{\rotatebox[origin=c]{90}{Actual}}} %
&0 & 44 & 31& 1 &  &   &   \\ 
&1 & 10 & 18& 2&  0&   & \\
&2 & 3 & 16& 11&  2&  1&  \\
&3 &   &  1& 17& 15&  6& 1\\
&4 &   &  1&   &  3&  5& 1\\
&5 &   &   &   &  1&  4& 1\\ 
\hline
\end{tabular}
\end{table}

The performance of the system can also be assessed with respect to the inter-reader variability. 
The VE scores were computed from the consensus reading of two research analysts, adjudicated by a third radiologist reading in the case of disagreements. 
This was assessed by comparing the inter-analyst Spearman's rank correlation of $\rho_{aa} =$ 0.84, 95\% CI [0.80 - 0.88] to the first-analyst system correlation of $\rho_{as} =$ 0.79, 95\% CI [0.74 - 0.84]. 
This gives $P(\rho_{aa} > \rho_{as}) =$ .93. 
The confidence intervals and probability are computed via bootstrapping with 10k samples.

VE scores were assessed based on the location in the scan with the highest visible grade of emphysema. Even if the emphysema is very localized, this can lead to a degree of inter-reader variability in the ground truth annotation. Given these challenging conditions, the high level of performance at the task is impressive. Furthermore the clinical relevance of the results should be emphasized. The VE score has independently associated with mortality \cite{Lync17}, indicating that this method can provide a similarly effective mortality signal. Additionally VE scores were associated with genetic markers linked to COPD susceptibility \cite{Halp17}. 

%The classifier can also be evaluated in relation to the reader performance.
%This was done by scoring scans for which there were multiple readings.
%The automatic scoring was treated like an additional observer for which the inter observer variability was computed. This gave the value XX which is YY compared to untrained readers and ZZ compared to trained readers.

\subsection{COPD Classification}

The next task was FEV$_1$/FVC prediction. 
In contrast to VE scoring, this quantity is defined via a lung function test. 
It is correlated to the lung appearance as can be seen in the correlation with the VE score as shown in \autoref{fig:fof_vs_cle}. 
The results from the automatic system can be seen in \autoref{fig:fof_computed}. 
From the slightly lower correlation with visual VE scores it can be seen that it is a more difficult for the system to predict this quantity visually.

\begin{figure}[ht!]
\centering
  \includegraphics[width=.50\linewidth]{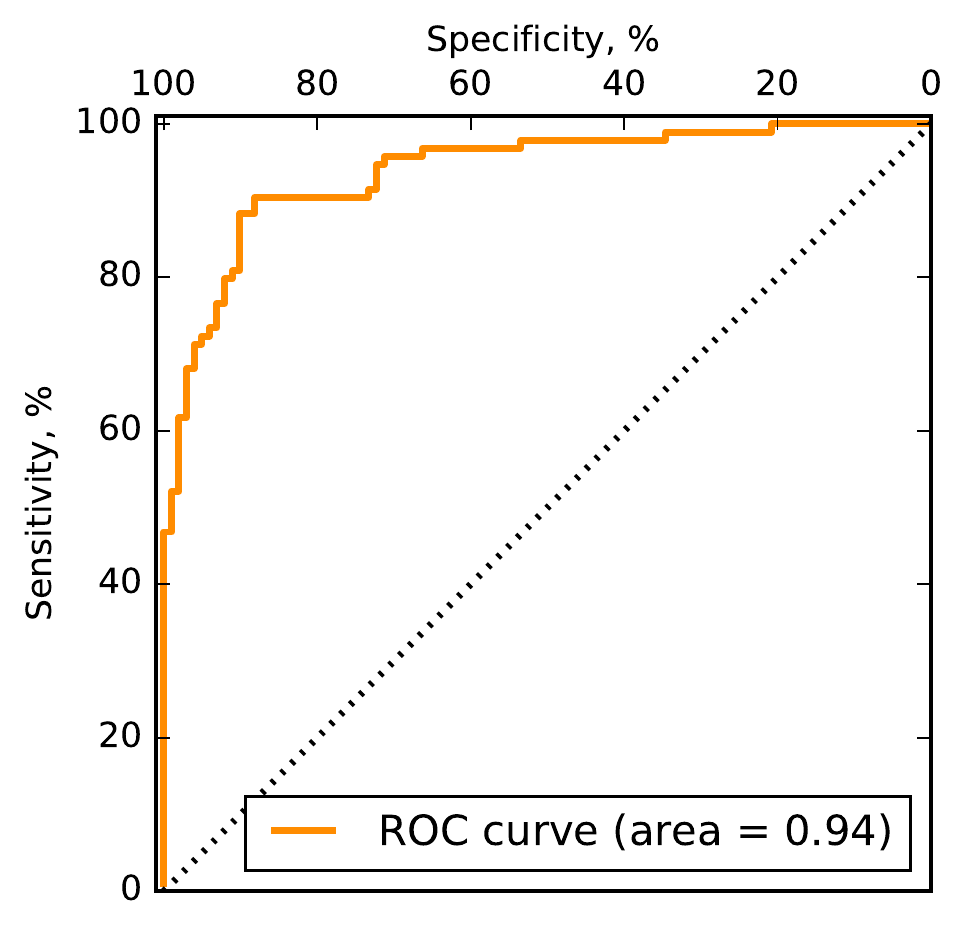}
  \caption{ROC curve for COPD classification using computed FEV$_1$/FVC values.}
  \label{fig:fof_roc}
\end{figure}

FEV$_1$/FVC is important because a patient is diagnosed with COPD if this value is below 0.7.
This discretization allows a confusion matrix, \autoref{tag:copd_cm}, and accuracy metrics to be computed.
Additionally a ROC curve can be computed by using the numerical value of FEV$_1$/FVC as a score as shown in \autoref{fig:fof_roc}.

This method outperforms other CT based COPD prediction methods shown in \autoref{tab:copd_comparison}, even without making use of additional patient information. 
For example age, body mass index, smoking status and pack-years of smoking history are used in \cite{Mets13b} together
with the CT derived measures of air trapping, for which an additional expiration scan is needed.

\vspace{1.5em}
\hspace{-.75em}\begin{minipage}{\textwidth}
  \begin{minipage}[b]{0.34\textwidth}
    \centering
      \begin{tabular}{| c | c c |}
      \hline
       COPD  & \multicolumn{2}{c|}{Predicted} \\
       Present & No & Yes \\
      \hline
      No  & 91 & 10   \\ 
      Yes & 11 & 83 \\ 
      \hline
      \end{tabular}
      \captionof{table}{  Confusion Matrix of COPD diagnosis comparing spirometric and computed values.   \label{tag:copd_cm}}
  \end{minipage}
  \hfill
  \hspace{1.5em}
  \begin{minipage}[b]{0.59\textwidth}
    \centering
\begin{tabular}{| l | c c |}
\hline
Method & AUC & 95\%CI \\
\hline
Mets\cite{Mets11a}             & 0.83 &0.81-0.86 \\
Mets\cite{Mets13b} Asymptomatic& 0.83 &0.80-0.87 \\
Mets\cite{Mets13b} Symptomatic & 0.91 &0.88-0.93 \\
\textbf{Ours} & 0.94 & 0.91-0.96\\
\hline
\end{tabular}
      \captionof{table}{    Performance comparison of COPD prediction methods. (Not evaluated on the same dataset.) \label{tab:copd_comparison} }
  \end{minipage}
\end{minipage}

\subsection{GOLD Stage Prediction}
The next quantity we looked at was computing FEV$_1$\%predicted. 
This is a normalized form of FEV$_1$ to account for variation in height and age.
Depending on the FEV$_1$\%predicted value patients have, they are stratified into different GOLD stages, which measure the severity of COPD.
This quantity is even less strongly correlated with VE score, see \autoref{fig:fev_vs_cle}, and correspondingly is harder to predict, see \autoref{fig:fev_computed}.
Again we follow the same evaluation procedure of discretization and accuracy evaluation.
\vspace{1.5em}

\begin{minipage}{\textwidth}
  \begin{minipage}[b]{0.48\textwidth}
    \centering
\begin{tabular}{|c c | c c c c c|}
\hline
 \multicolumn{2}{|c|}{GOLD}  & \multicolumn{5}{c|}{Predicted} \\
 \multicolumn{2}{|c|}{Stage} & 0 & 1 &	2 & 3 & 4 \\
\hline
\parbox[t]{2mm}{\multirow{5}{*}{\rotatebox[origin=c]{90}{Actual}}}%
&0 & 91&  3&  7&   &   \\ 
&1 &  4&  6& 11&   &   \\
&2 &  6&  5& 28&  5&   \\
&3 &  1&   & 13&  8& 1 \\
&4 &   &   &  1&  4& 1 \\
\hline
\end{tabular}
\captionof{table}{ Confusion Matrix of Gold Stages. \label{tag:gold_cm} }
  \end{minipage}
  \hfill
  \begin{minipage}[b]{0.48\textwidth}
    \centering
\begin{tabular}{| c | c c c |}
\hline
num. slices & $\rho$ & \multicolumn{2}{c|}{95\% CI} \\
\hline
8  & 0.78 & 0.72 & 0.82 \\
12 & 0.79 & 0.74 & 0.84 \\
16 & 0.82 & 0.77 & 0.86 \\
24 & 0.81 & 0.76 & 0.85 \\
\hline
\end{tabular}
\captionof{table}{ Model Variation trained on FEV$_1$/FVC. \label{tag:model_variation} }
\end{minipage}
\end{minipage}

\subsection{Model Search}

One important hyper-parameter of the model is the number of slices that are samples from the scan.
In order to show how performance depends on this parameter, we tested how well FEV$_1$/FVC could be computed with a varying number of slices.
The result of this can be seen in \autoref{tag:model_variation}.

\begin{table}[!ht]
\centering
\end{table}

\section{Discussion}

We have shown that a multi-slice analysis of full 3D CT scans, trained end-to-end, can make meaningful predictions at the patient level. Exactly the same approach can be used to predict multiple outputs, without the need to encode any specific knowledge into the system about which characteristics on the scan are known to be correlated to the output parameter that we want to predict. This is in contrast to previously published methods to predict the severity of parenchymal emphysema, lung function status and COPD status. These works were based on the detection and quantification of specific local patterns in the scan, derived from lung parenchyma and the airways. One would expect that such dedicated methods would outperform the generic approach proposed here. The proposed approach could function as a baseline to benchmark specific approaches. But in fact our experimental results suggest that our system is at least equivalent or even superior to dedicated approaches. Further evaluation on common datasets is needed to analyze the full potential of deep convolutional methods versus dedicated classical approaches. 

There are various strategies that could be explored to maximize performance. The most obvious suggestion is to train the systems on a larger set of CT  scans. Especially for predicting lung function status, a much larger number of CTs with corresponding lung function tests are already contained in the COPDGene database. In this work we limited ourselves to the scans we could obtain from the COPDGene study for which all three output parameters we wanted to predict were available. This allowed us to perform all experiments with the same training set.

Another option is to add additional types of input data. Previous work focused on predicting COPD status included additional parameters about the patient, such as age, smoking status and history, and additional exposures are known to correlate with COPD. Such information can be inserted at various locations into the network to ensure that the entire system can still be trained end-to-end. Additionally, expiration CT scans could be added as an additional input, and the system would analyze slices from multiple scans. This approach can also be used to extend the analysis to longitudinal scan data. If data at multiple time points would be available, prediction of clinically relevant trends, e.g.\ the rate of COPD progression, can be expected to be more accurate. 

Slice sampling could be improved by evaluating several samples per scan and averaging the results. In the presented results, the system inspects only 16 slices from a 3D scan. One could offset the location of the start slice and run the same analysis multiple times. Furthermore, additional networks could be trained for coronal and sagittal slices and a combination of all three could be used to provide the final score.
This would go some way to address a major drawback of \scantonum, namely that it uses a 2D approach to analyze 3D data, and only analyzes a limited number of slices. Note that we have trained a system with 24 slices and achieved no improvement. We have also varied other aspects of the architecture. In particular, we investigated the use a Soft-Max loss for VE scoring, but this did not yield good results. Using L2 regression allowed the network to perform better. One could attempt to improve performance further in the case of tasks that produce discretized quantities by additional calibration that may compensate for prediction bias.

\begin{figure*}[!ht]
\centering
\includegraphics[width=.7\textwidth]{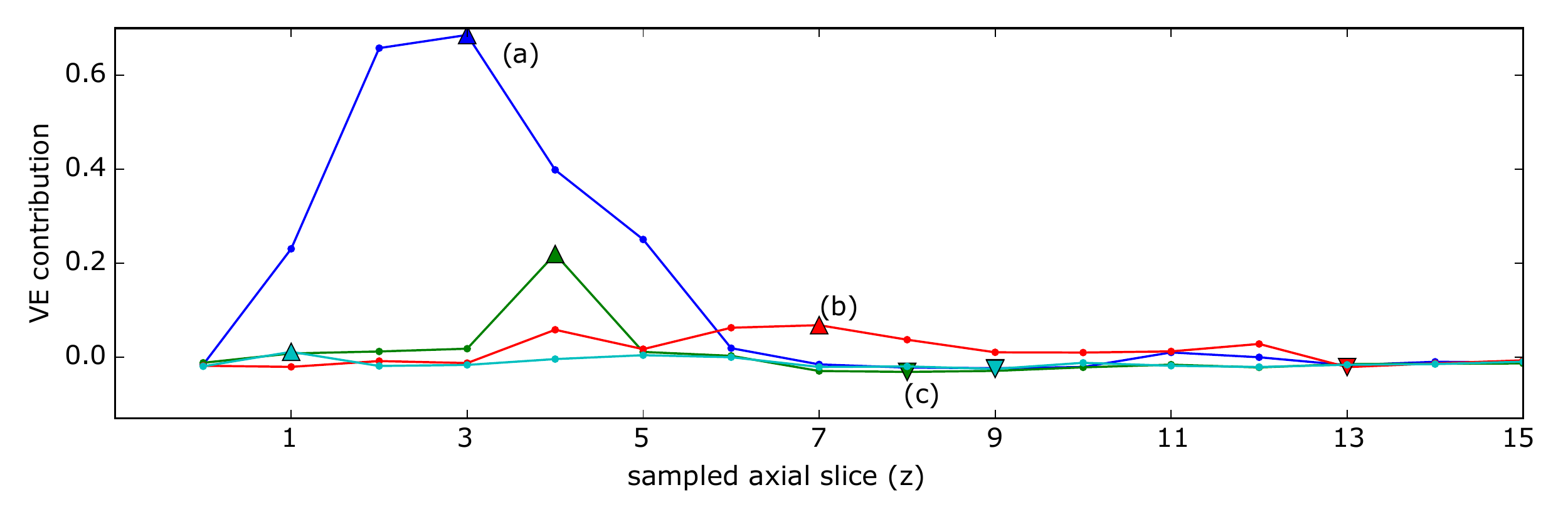}

\begin{tabular}{ccc}
%\hline
\includegraphics[width=.23\textwidth]{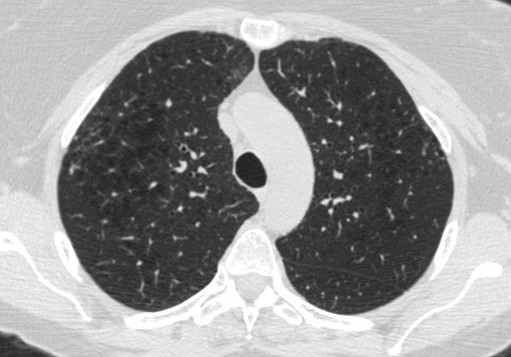}&%
\includegraphics[width=.23\textwidth]{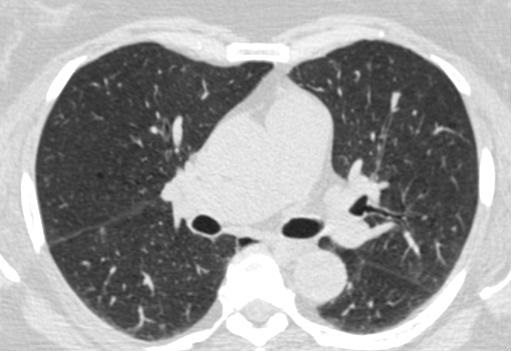}&%
\includegraphics[width=.23\textwidth]{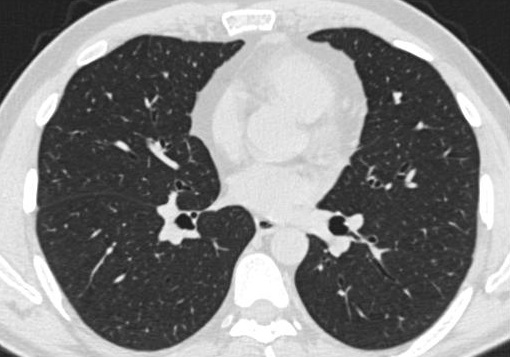}\\
(a)&(b)&(c)\\
\end{tabular}
\caption{The plot above shows VE score contributions for each sampled axial slices for four random scans. The maximal and minimal VE response are indicated by triangles, those marked with letters are shown below. Slice a shows extensive emphysema in the center of the right lung, scan b shows a small emphysema lesion above the right fissure and slice c shows healthy slices are scored correctly. The images are best viewed zoomed in on a large computer monitor.
\label{fig:extrema}}
\end{figure*}

% transposed version of extrema plot
\iffalse
\begin{figure*}
\centering
\begin{tabular}{m{4cm}m{1cm}l}
%\hline
\vspace{-5cm}%
\multirow{3}{0cm}{\centering \includegraphics[width=4cm,height=9cm]{extrema/diff2.pdf}}%&&\\
%\hline
&\vspace{-3cm}(a),(b)&\includegraphics[width=3cm,height=3cm]{extrema/extrema_a.jpg}%
\includegraphics[width=3cm,height=3cm]{extrema/extrema_b.jpg}\\
%\hline
&\vspace{-3cm}(c), (d)&\includegraphics[width=3cm,height=3cm]{extrema/extrema_c.jpg}%
\includegraphics[width=3cm,height=3cm]{extrema/extrema_d.jpg}\\
%\hline
&\vspace{-3cm}(e), (f)&\includegraphics[width=3cm,height=3cm]{extrema/extrema_e.jpg}%
\includegraphics[width=3cm,height=3cm]{extrema/extrema_f.jpg}\\
%\hline
\end{tabular}
\caption{VE score contributions of individual slices shown non the left, with network input images shown for extreme values of VE score, shown right.
\label{fig:extrema}}
\end{figure*}
\fi

A potential drawback of the overall approach, that may limit wider applicability to other tasks, is the requirement of a large set of labeled scans. For this work the COPDGene trial provided these, making it an excellent setting to explore the performance of such architectures.
We hope that these datasets can be used as benchmarks for evaluating architecture improvements such as those mentioned above. 
Better performance would allow the use of fewer scans, enabling the application of this approach to a larger set of problems. \note{can you say anything specific how you'd do that? -> Max: I reformulated it.}
%Thee varying degree of localization between CLE and PSE can offer additional insights into the trade-off of scope.
%While we make use of some methods to mitigate these drawbacks such as data augmentation, network architecture, slices)

%Comparison of performance with other papers is
%Especially when comparing quality metrics of discretized quantities care should be taken to have %a similar distribution of samples.
%For example we would expect that if an evaluation dataset had more evenly distributed frequencies of emphysema severity or GOLD stage, the accuracy would be increased, because the natural frequency dataset is biased towards low GOLD classes which are more difficult to distinguish.

%One of the main aims in COPD research is to try to identify rapidly progressing cohorts. This was also attempted but did not succeed.

In this work we have predicted lung function tests results from CT scans. Obviously performing lung function tests directly is a viable alternative. We like to stress that in clinical practice an enormous amount of chest CT scans are acquired in subjects who are at risk for chronic lung diseases but do not undergo lung functions testing. This is especially true now CT lung cancer screening programs are being implemented. CT based surrogate lung function measurements may therefore be a good and extremely cheap way to also screen for COPD, as suggested by Mets et al.~\cite{Mets11a}. If visual surrogates become better, they could be used as a second measurement for lung function in order to decrease the variance of this test. It may be possible to use similar methods to identify which lung cancer screening participants are particularly at risk of the disease and to adjust screening frequency or follow up accordingly. A recent study indicated that risk for cardiovascular disease can be assessed by visually by radiologists \cite{Chil15}, possibly \scantonum could achieve similar results for this task.

\section{Conclusion}

This paper shows that learning VE, FEV$_1$/FVC and FEV$_1$ from whole scans is possible, even with relatively simple networks.
Whole scan end-to-end learning is a general method which can be applied to a large number of diagnostic and image quantification processes in radiology.
We hope to have shown that the proposed method is sufficiently attractive for it to be considered for other tasks.

\section*{Acknowledgments}

This work was supported in part by the VICI project 016.130.326 of the Dutch Science Foundation (NWO). We gratefully acknowledge the COPDGene Study (ancillary study ANC-251) for providing the data used.

% Either type in your references using
% \begin{thebibliography}{}
% Text
% \end{thebibliography}
%
% or
%
% Compile your BiBTeX database using our plos2015.bst
% style file and paste the contents of your .bbl file
% here. See http://journals.plos.org/plosone/s/latex for 
%\ifCLASSOPTIONcaptionsoff
%  \newpage
%\fi


\bibliographystyle{IEEEtran}
\bibliography{fullstrings,IEEEabrv,copd}

\begin{thebibliography}{10}

\bibitem{Lync15}
Lynch DA, Austin JHM, Hogg JC, Grenier PA, Kauczor HU, Bankier AA, et~al.
\newblock {CT}-Definable Subtypes of {C}hronic {O}bstructive {P}ulmonary
  {D}isease: A Statement of the {F}leischner {S}ociety.
\newblock Radiology. 2015;277(1):192--205.
\newblock doi:{10.1148/radiol.2015141579}.

\bibitem{Lync12}
Lynch DA, Murphy JR, Crapo JD, Criner GJ, Galperin-Aizenberg M, Jacobson FL,
  et~al.
\newblock A Combined Pulmonary -Radiology Workshop for Visual Evaluation of
  {COPD}: Study Design, Chest {CT} Findings and Concordance with Quantitative
  Evaluation.
\newblock COPD. 2012;9.2:151 -- 159.
\newblock doi:{10.3109/15412555.2012.654923}.

\bibitem{Mosl16}
Mosley J, Smith L, Dutton B.
\newblock {Tiotropium Bromide/Olodaterol (Stiolto Respimat)}: Once-Daily
  Combination Therapy for the Maintenance of {COPD}.
\newblock Pharmacy and Therapeutics. 2016;41(2):97--102.

\bibitem{Male14}
Maleki-Yazdi MR, Kaelin T, Richard N, Zvarich M, Church A.
\newblock Efficacy and safety of umeclidinium/vilanterol 62.5/25 mcg and
  tiotropium 18 mcg in chronic obstructive pulmonary disease: Results of a
  24-week, randomized, controlled trial.
\newblock Respiratory Medicine. 2014;108(12):1752 -- 1760.
\newblock doi:{http://dx.doi.org/10.1016/j.rmed.2014.10.002}.

\bibitem{Rikx09}
van Rikxoort EM, de~Hoop B, van~de Vorst S, Prokop M, van Ginneken B.
\newblock Automatic segmentation of pulmonary segments from volumetric chest
  {CT} scans.
\newblock IEEE Transactions on Medical Imaging. 2009;28:621--630.
\newblock doi:{10.1109/TMI.2008.2008968}.

\bibitem{Han11}
Han MK, Kazerooni EA, Lynch DA, Liu LX, Murray S, Curtis JL, et~al.
\newblock Chronic Obstructive Pulmonary Disease Exacerbations in the COPDGene
  Study: Associated Radiologic Phenotypes.
\newblock Radiology. 2011;261(1):274--282.
\newblock doi:{10.1148/radiol.11110173}.

\bibitem{Stoe08}
Stoel BC, Putter H, Bakker ME, Dirksen A, Stockley RA, Piitulainen E, et~al.
\newblock Volume Correction in Computed Tomography Densitometry for Follow-up
  Studies on Pulmonary Emphysema.
\newblock Proceedings of the American Thoracic Society. 2008;5(9):919--924.
\newblock doi:{10.1513/pats.200804-040QC}.

\bibitem{Gall16}
Gallardo-Estrella L, Lynch DA, Prokop M, Stinson D, Zach J, Judy PF, et~al.
\newblock Normalizing computed tomography data reconstructed with different
  filter kernels: effect on emphysema quantification.
\newblock European Radiology. 2016;26:478--486.
\newblock doi:{10.1007/s00330-015-3824-y}.

\bibitem{Xu06}
Xu Y, Sonka M, McLennan G, Guo J, Hoffman EA.
\newblock {MDCT-based 3-D} texture classification of emphysema and early
  smoking related lung pathologies.
\newblock IEEE Transactions on Medical Imaging. 2006;25(4):464--475.
\newblock doi:{10.1109/TMI.2006.870889}.

\bibitem{Park08}
Park YS, Seo JB, Kim N, Chae EJ, Oh YM, Lee SD, et~al.
\newblock Texture-Based Quantification of Pulmonary Emphysema on
  High-Resolution Computed Tomography: Comparison With Density-Based
  Quantification and Correlation With Pulmonary Function Test.
\newblock Investigative Radiology. 2008;43(6):395--402.
\newblock doi:{10.1097/RLI.0b013e31816901c7}.

\bibitem{Sore10}
S{\o}rensen L, Shaker SB, de~Bruijne M.
\newblock Quantitative Analysis of Pulmonary Emphysema Using Local Binary
  Patterns.
\newblock IEEE Transactions on Medical Imaging. 2010;29(2):559--569.
\newblock doi:{10.1109/TMI.2009.2038575}.

\bibitem{Haem14}
H\"{a}me Y, Angelini ED, Hoffman EA, Barr RG, Laine AF.
\newblock Adaptive Quantification and Longitudinal Analysis of Pulmonary
  Emphysema With a Hidden {M}arkov Measure Field Model.
\newblock IEEE Transactions on Medical Imaging. 2014;33(7):1527--1540.
\newblock doi:{10.1109/TMI.2014.2317520}.

\bibitem{Yang16}
Yang J, Feng X, Angelini ED, Laine AF.
\newblock Texton and sparse representation based texture classification of lung
  parenchyma in {CT} images.
\newblock In: 2016 38th Annual International Conference of the IEEE Engineering
  in Medicine and Biology Society (EMBC); 2016. p. 1276--1279.

\bibitem{Gins16}
Ginsburg SB, Zhao J, Humphries S, Jou S, Yagihashi K, Lynch DA, et~al.
\newblock Texture-based Quantification of Centrilobular Emphysema and
  Centrilobular Nodularity in Longitudinal {CT} Scans of Current and Former
  Smokers.
\newblock Academic Radiology. 2016;23(11):1349 -- 1358.
\newblock doi:{http://dx.doi.org/10.1016/j.acra.2016.06.002}.

\bibitem{Bind16}
Binder P, Batmanghelich NK, Estepar RSJ, Golland P.
\newblock Unsupervised Discovery of Emphysema Subtypes in a Large Clinical
  Cohort.
\newblock In: Wang L, Adeli E, Wang Q, Shi Y, Suk HI, editors. Machine Learning
  in Medical Imaging: 7th International Workshop, MLMI 2016, Held in
  Conjunction with MICCAI 2016, Athens, Greece, October 17, 2016, Proceedings.
  Cham: Springer International Publishing; 2016.

\bibitem{Chep14}
Cheplygina V, S{\o}rensen L, Tax DMJ, Pedersen JH, Loog M, d~Bruijne M.
\newblock Classification of {COPD} with Multiple Instance Learning.
\newblock In: 2014 22nd International Conference on Pattern Recognition; 2014.

\bibitem{Mets12}
Mets OM, Murphy K, Zanen P, Gietema HA, Lammers JW, van Ginneken B, et~al.
\newblock The relationship between lung function impairment and quantitative
  computed tomography in chronic obstructive pulmonary disease.
\newblock European Radiology. 2012;22:120--128.
\newblock doi:{10.1007/s00330-011-2237-9}.

\bibitem{Oakd17}
Oakden-Rayner L, Carneiro G, Bessen T, Nascimento JC, Bradley AP, Palmer LJ.
\newblock Precision Radiology: Predicting longevity using feature engineering
  and deep learning methods in a radiomics framework.
\newblock Scientific Reports. 2017;7(1648).

\bibitem{Rega10}
Regan EA, Hokanson JE, Murphy JR, Make B, Lynch DA, Beaty TH, et~al.
\newblock Genetic Epidemiology of {COPD (COPDGene)} Study Design.
\newblock COPD. 2010; p. 32--43.
\newblock doi:{10.3109/15412550903499522}.

\bibitem{Halp17}
Halper-Stromberg E, Cho MH, Wilson C, Nevrekar D, Crapo JD, Washko G, et~al.
\newblock Visual Assessment of Chest Computed Tomographic Images Is
  Independently Useful for Genetic Association Analysis in Studies of Chronic
  Obstructive Pulmonary Disease.
\newblock Annals of the American Thoracic Society. 2017;14(1):33--40.
\newblock doi:{10.1513/AnnalsATS.201606-427OC}.

\bibitem{Karp14}
Karpathy A, Toderici G, Shetty S, Leung T, Sukthankar R, Fei-Fei L.
\newblock Large-scale Video Classification with Convolutional Neural Networks.
\newblock In: CVPR; 2014.

\bibitem{Slui05a}
Sluimer IC, Prokop M, van Ginneken B.
\newblock Towards automated segmentation of the pathological lung in {CT}.
\newblock IEEE Transactions on Medical Imaging. 2005;24:1025--1038.
\newblock doi:{10.1109/TMI.2005.851757}.

\bibitem{Glor10}
Glorot X, Bengio Y.
\newblock Understanding the difficulty of training deep feedforward neural
  networks.
\newblock In: In Proceedings of the International Conference on Artificial
  Intelligence and Statistics (AISTATS'10). Society for Artificial Intelligence
  and Statistics; 2010.

\bibitem{jia14}
Jia Y, Shelhamer E, Donahue J, Karayev S, Long J, Girshick R, et~al.
\newblock Caffe: Convolutional Architecture for Fast Feature Embedding.
\newblock arXiv preprint arXiv:14085093. 2014;.

\bibitem{Lync17}
Lynch DA, Moore C, Wilson CG, Nevrekar DV, Jennermann TB, Humphries S.
\newblock Visual Emphysema Pattern Using the {Fleischner Society}
  Classification System Is Independently Associated with Mortality in Cigarette
  Smokers.
\newblock In: Annual Meeting of the Radiological Society of North America;
  2017.

\bibitem{Mets13b}
Mets OM, Schmidt M, Buckens CF, Gondrie MJ, Isgum I, Oudkerk M, et~al.
\newblock Diagnosis of chronic obstructive pulmonary disease in lung cancer
  screening Computed Tomography scans: independent contribution of emphysema,
  air trapping and bronchial wall thickening.
\newblock Respiratory Research. 2013;14:59.
\newblock doi:{10.1186/1465-9921-14-59}.

\bibitem{Mets11a}
Mets OM, Buckens CFM, Zanen P, Isgum I, van Ginneken B, Prokop M, et~al.
\newblock Identification of Chronic Obstructive Pulmonary Disease in Lung
  Cancer Screening Computed Tomographic Scans.
\newblock Journal of the American Medical Association. 2011;306:1775--1781.
\newblock doi:{10.1001/jama.2011.1531}.

\bibitem{Chil15}
Chiles C, Duan F, Gladish GW, Ravenel JG, Baginski SG, Snyder BS, et~al.
\newblock Association of Coronary Artery Calcification and Mortality in the
  {N}ational {L}ung {S}creening {T}rial: A Comparison of Three Scoring Methods.
\newblock Radiology. 2015;276:82--90.
\newblock doi:{10.1148/radiol.15142062}.

\end{thebibliography}


%\vfill


\end{document}